# Generalized network recovery based on topology and optimization for real-world systems

**Authors:** Udit Bhatia[1], Lina Sela Perelman[2], Auroop Ratan Ganguly[1]*

**Affiliations:**

[1]Civil and Environmental Engineering, Northeastern University, Boston, Massachusetts, United States, 02115

[2]Civil, Architectural and Environmental Engineering, University of Texas at Austin, Texas, United States, 78712.

*__Correspondence to__: Auroop Ratan Ganguly, Sustainability and Data Sciences Lab, Department of Civil & Environmental Engineering, 400 Snell Engineering, Northeastern University, 360 Huntington Avenue, Boston, MA 02115, USA; Phone: +1-617-373-6005; Email: a.ganguly@neu.edu

**Abstract:** Designing effective recovery strategies for damaged networked systems is critical to the resilience of built, human and natural systems. However, progress has been limited by the inability to bring together distinct philosophies, such as complex network topology through centrality measures and network flow optimization through entropy measures. Network centrality-based metrics are relatively more intuitive and computationally efficient while optimization-based approaches are more amenable to dynamic adjustments. Here we show, with case studies in real-world transportation systems, that the two distinct network philosophies can be blended to form a hybrid recovery strategy that is more effective than either, with the relative performance depending on aggregate network attributes. Direct applications include disaster management and climate adaptation sciences, where recovery of lifeline networks can save lives and economies.

**One Sentence Summary:** A hybrid method combining topology and optimization for recovering damaged networks demonstrated on real-world transportation systems

Lifelines provide essential services to residents and businesses across geographic scales and help ensure the public's health and safety, as well as economic security (*1*). Energy, water, transportation, and communications represent the four primary lifelines, and comprise interdependent networked systems, such as the infrastructures that support power grids, water distribution or wastewater systems, multi-modal transportation including railways, roads, airways, or waterways, and telephone, satellite, or Internet services. Natural, technological, and man-made catastrophes, including weather extremes and cyber-physical attacks, may severely disrupt the functioning of these lifeline infrastructure networks, thus causing loss of essential services. The corresponding impacts can be felt by rural and urban communities with cascading fallouts across cities, megalopolises, nations, regions, and indeed the entire globe (*2*). The ability to manage, adapt to, and mitigate these "globally networked risks" (*3*, *4*) may determine the extent to which human society can benefit from globalization versus becoming increasingly vulnerable to large-scale failures. Probabilistic risk analysis, reliability engineering, operations research, and in recent years, network science (*5–9*) have demonstrated value in characterizing, designing, maintaining, and operating Lifeline Infrastructure Networked Systems (LINS). However, once disasters strike, intervention by emergency managers, stakeholders, and policymakers is imperative for the timely and efficient restoration of the essential services supported by LINS. One crucial knowledge gap in this context is the inability to provide systematic guidance for post-disruption recovery, let alone prepare such recovery plans in advance. While optimization approaches have demonstrated value in simulated or stylized settings and network science methods have shown initial promise with simulated or real-world networks (*5*, *7*, *10–12*), principled strategies for recovery, especially those that can benefit from both, are still lacking.

Current efforts in infrastructure recovery (*5*, *13*, *14*) include the post-disaster maximum flow, connectivity recovery, identification of key components for resource allocation using network science-based approaches, and theoretical recovery frameworks for simulated networks (*7*, *10*, *13*). Existing infrastructure recovery methods either focus on network heuristics that prioritize restoration based on pre-defined measures, or optimization methods which have been tested either on stylized or simulated networks (*11*). While centrality-based metrics tend to be intuitive and computationally efficient but remain static irrespective of the desired essential functionality of networked infrastructures. Optimization-based approaches, while usually less intuitive and more computationally expensive, can potentially be better tailored to design objectives.

However, most real-world networks often differ from the stylized models and assessing the impact of disruptive events followed by recovery calls for a case-by-case assessment of the most efficient recovery strategies. For example, a recent study proposed a methodology for joint restoration modeling of interdependent infrastructure systems exemplified through interdependent gas and power system at county level impacted by a hurricane. Here, genetic algorithm-derived scores were used to generate recovery sequences. As noted by the researchers, the proposed framework requires multiple models to function, including a hazard generation model, component fragility models and system performance models (*9*). We note that in this framework, damage and recovery of components (or nodes) was informed by disaster-specific fragility curves, which would not be readily available for many infrastructure systems which operate at large geographical scales and face a variety of potential threats (unlike smaller urban or county scale networks) (*15*).

To address the shortcomings, we develop an edge recovery algorithm that blends network science-based heuristics such as network centrality measures and optimization approaches such a greedy algorithm and cross entropy algorithm to inform recovery of real-world networks after disparate hazards. This approach enables us to design dynamic objective functions in which real-world constraints such as resources, recovery time, a combination of nodes and edge failure can be included. We apply the model to two real-world transportation systems: Indian Railways Network, and Boston's Mass Transit System (MBTA) (*16–18*). We model network disruptions inspired by real-life events. Then, we apply the proposed edge recovery algorithms to compare performance of various strategies. The results may reveal important considerations for assessing proposals for disaster preparedness that in turn can help stakeholders to act in advance to restore network functionality at faster rates.

**Critical functionality of lifelines**

In the context of lifelines, critical functions are defined as those functions that, if perturbed, can lead to serious societal emergency and crisis. Post-hazard restoration efforts ought to restore these systems to the level of initial performance. Hence, we use State of Critical Functionality (SCF) as a performance measure, which is defined as a ratio of functionality at a given instance during restoration process (Instantaneous Functionality or IF) to a state of functionality before perturbation (Target Functionality or TF). Hence, SCF of 1 represents a fully functional system, whereas SCF equal to zero represents the total loss of functionality. In transportation context, the impact of perturbations unfolds in form of loss or impairment of stations and/or linkages between these stations which in turn results in loss of critical functions such as delays, disruption of traffic flow and loss of connectivity. In the present study, we use two different measures for SCF for the systems under consideration: (a) network connectivity (assessed by measuring the size of the Largest Connected Component (LCC) in a network) (b) satisfying origin-destination demand (measured as a total traffic flow in LCC).

**Perturbation and Recovery**

To simulate perturbations on IRN, we consider three hazards that are inspired by real-life events: (a) simulation inspired by 2004 Indian Ocean tsunami which impacted stations and tracks on eastern coast of India; (b) scenario based on a cascade from the power grid, based on the historically massive 2012 blackout; and (c) a Cyber or Cyber-physical attack scenario, where stations are maliciously targeted based on traffic volume. We call these events as Tsunami, Grid and Cyber-physical, respectively in future sections for brevity.

Consequences of Cyber or Cyber-Physical attacks are based on hypothetical scenarios, although motivated from real-world events such as 26/11 Mumbai terror attack. For MBTA, we consider a hazard that is inspired by Nor'easter of 2015 (*19*). These hazards result in loss of impacted nodes and edges which are incident on these nodes. Loss of these components, in turn, results in reduced functionality as the system cannot sustain the TF. (See table S1 and table S2 for the list of affected stations by various hazards in the two networks).

Post-perturbation system recovery entails strategic restoration of constituting components of the network that are impacted by perturbation. When multiple components are lost, determining recovery sequences for restoration plays a crucial role in regaining intended performance levels for the system (*13*). In this study, we focus on determining recovery sequence of edges which results in a fast recovery to the pre-disaster functionality, i.e. minimize resilience loss (Fig. 2).

Recovery takes on the following process:

1. SCF is computed at the initial post-hazard state.
2. Prioritization sequence is identified. This prioritization sequence is the order in which edges should be restored.
3. Instead of directly restoring nodes, we focus on restoring edges and node is said to be restored once it becomes part of the largest connected component of the network. The sequence can be generated through network science metrics or optimization approaches (discussed later in this section).
4. Given a sequence, iteratively restore edges until SCF =1:
a. The next edge in the prioritization sequence is restored, establishing flow between a pair of nodes that it is connecting. This grows the size of the largest connected component and enables traffic flow across the network.
b. FF is recalculated.
c. SCF is recalculated.

In this study, we build on Network Science based centrality measure and optimization approaches for network recovery to generate recovery sequences after each perturbation simulated on the two systems under consideration (See Methods).

**Results**

We measure efficiency of each recovery strategy by computing the corresponding resilience loss, which is defined as area between Y-axis and recovery curve (Fig. 2). Lower area means higher efficiency as it represents the amount of resources required to bounce back to the functionality at which system was operating before perturbation.

We use connectivity and traffic volume as the two performance measures for State of Critical Functionality (SCF). We simulate various hazard scenarios on the two networks under consideration (See materials and methods). For IRN, SCF dropped by 14%, and 16%; cascades of failure from the power grid, based on 2012 blackout resulted in SCF loss of 30% and 35%; whereas simulated Cyber-physical attack ensued loss of 27% and 35% of SCF in terms of connectivity and traffic volume, respectively. To restore the SCF of perturbed network to Targeted Functionality (TF), we use edge prioritization sequences determined from network science based and optimization-based approaches. We find that for IRN, optimization-based results in better performance than network science-based approaches. On an average, optimization strategies are 12% and 40% more efficient when compared with centrality-based strategies for both connectivity recovery and traffic flow recovery, respectively (Fig. 3 and Table 1).

For MBTA, simulation of 2015 Nor'easter inspired exhibit disproportionately adverse impacts on SCF as removal of 9% edges (17/176) resulted in 80% loss of SCF in terms of connectivity as well as traffic volume. Contrary to our finding for IRN, we observe that network science-based recovery approach, specifically edge betweenness centrality, outperforms optimization-based GA approach by 41% whereas efficiency from CE was comparable to the edge betweenness centrality (Fig. 3).

On the other hand, we find that for MBTA, which approaches tree-like network topology, approaches such as GA that tend to maximize instant or short-term restoration of functionality may turn out to be myopic and inefficient at system level over the lifetime of recovery process (Fig. 4). Specifically, Network-based recovery approach outperform optimization based Greedy Approach by 41% and 38% for network connectivity recovery and traffic flow recovery, respectively. Edge centrality-based recovery marginally better than CE. Both CE and edge centrality outperforms GA by 40%. (Table 2). This holds true for both connectivity recovery and traffic flow recovery.

To ensure that our results are robust to the choice of parameters, we perform CE optimization using 3 different choices of parameters and we observe that efficiency scores thus obtained are insensitive to parameter choice for both objective functions (Fig. S1). To understand the relationship between topological features of the network and recovery scores, we plot degree distribution and weighted degree distribution of the network. In addition, we also measure network assortativity to measure the similarity of connections in the network with respect to node degree. We find that Indian Railways Network follows the power law distribution in degree as well as weighted degree distribution (Fig. 5(A)). That is, degree distribution follows the form:

$$P(k) \sim k^{-\gamma} \quad [1]$$

where fraction $P(k)$ is the fraction of nodes in the network which have k connections, and gamma = 1.84 (p-value <0.05: student-T test). As shown in our prior research (*7, 10*), for networks exhibiting exponential decay in their degree distribution, centrality measures are often correlated with degree of nodes and hence generating node restoration sequences using network centrality attributes often yield high recovery efficiency. However, this intuition may not hold true in case of edge recovery. We find that degree assortativity coefficient for this network is 0.062 which means that there is no preference of a network node of high degree to attach to other high degree nodes (*20*). As a result, ranking edges based on the pair of attributes of nodes does not yield most efficient recovery trajectories. Boston's MBTA, on other hand, exhibit tree-like structure (*21*) with 80% of nodes having degrees in the range of 1 and 3 (Fig. 5(B)). For localized damage on tree-like networks such as MBTA, GA approach fails to identify the system-wide optimal just by maximizing functionality at each step because attributes of edges in tree-like network are similar to each other. Edges with higher edge betweenness (*22*) tend to "bridge" the broken network at a faster rate, therefore, exhibiting high efficiency. Recovery scores from various recovery strategies for both networks are summarized in Table 1.

**Discussion**

Post-perturbation network recovery strategies have been motivated by two distinct philosophies, specifically, centrality measures in complex networks versus network optimization including entropy measures. Here we show that they may be blended to offer complementary value for the recovery of networked infrastructure systems (Fig.3-4; Table 1). We design a strategy that blends network science and optimization to improve network recovery post-perturbation and demonstrate on two real-work networks, specifically, the Indian Railways Network (IRN) and the Massachusetts Bay Transportation Authority (MBTA) network. Our hybrid approach shows that the optimal algorithm at each recovery step may be situation-specific and allowing an automated way to choose between network science versus network optimization methods may result in gains of efficiency by anywhere from about 10% to about 40% for both network connectivity and traffic flow recovery. Furthermore, the performance can be mapped to the network attributes. Thus, optimization approaches work better for the IRN, which approaches scale-free network attributes, while network-centrality approaches work better for MBTA, which approaches a planar network. However, the performance at any specific recovery step may be dependent on the characteristics of the current attribute of the damaged network. Our analysis offers several meaningful inferences to be made that may have important implications for resilience policy for systems operating at various jurisdictional and geographical levels. Approaches that tend to maximize instant or short-term restoration of functionality may turn out to be myopic and inefficient at system level over the lifetime of recovery process.

Recovery strategies may need to handle tradeoffs between multiple and potentially disparate essential services which they were designed to provide. Recovery approaches need to adjust to current state of a lifeline infrastructure network rather than being exclusively guided by the topology of a system prior to loss of functionality. The network topology and flow attributes of a specific lifeline system determines the extent to which systematic or dynamics approaches may improve recovery of any specific lifeline network or generalize to other lifelines. Recovery strategies that dynamically consider multiple essential services and account for both network and flow attributes tend to be more reliable and generalizes better across systems. Infrastructure and lifeline recovery strategies have traditionally tended to be bottom-up where component specific and granular information are used where available, in the absence of which relatively ad hoc strategies become the default choice. The top-down approaches proposed here offers a way for infrastructure owners and operators to make recovery strategies that may be optimal at overall system level functionality.

Our hybrid approach shows that the optimal algorithm at each recovery step may be situation-specific and allowing an automated way to choose between network science versus network optimization methods may result in gains of efficiency by anywhere from about 10% to about 40% for both network connectivity and traffic flow recovery. Specifically, we note the following:

1. IRN: Optimization based approaches are 9% and 25% more efficient than network centrality-based approaches for network connectivity recovery and traffic flow recovery, respectively. Optimization based approaches (both Greedy Approach and Cross-Entropy) are better at all recovery steps as well are more efficient in terms of resilience loss (defined as area between Y-axis and recovery curve). This holds true for all three scenarios: tsunami, power-grid cascade, and simulated cyber-physical attack.
2. MBTA: Network-based recovery approach outperform optimization based Greedy Approach by 41% and 38% for network connectivity recovery and traffic flow recovery, respectively. Edge centrality-based recovery marginally better than CE. Both CE and edge centrality outperforms GA by 40%. This holds true for both connectivity recovery and traffic flow recovery.
3. Furthermore, the performance can be broadly mapped to the network attributes. Thus, optimization approaches work better for the IRN, which exhibits exponentially decaying degree distribution (with presence of hubs), while network-centrality approaches work better for MBTA, which approaches a planar network. However, the performance at any specific recovery step may be dependent on the characteristics of the current attribute of the damaged network. Thus, the hybrid approach which selects the best performing approach at the recovery time steps would appear to be the best suited.

While future research may need to balance the traditional bottom-up approaches with emerging top-down methodologies, none of these address data limitations and information gaps on their own. Thus, the proposed approaches need to make assumptions about resource constraints, recovery time and fragility at component levels. However, while these assumptions do limit the validity of the conclusions in data limited study, they also point to what new data may add value to the analysis and help improve credibility of the results. Thus, infrastructure owners and operators may have direct access to component-specific data about resources and vulnerabilities which could augment the methodologies proposed and demonstrated here.

One final caveat is that the optimization algorithms adopted for network recovery in this work are known to be computationally expensive. This may require further research in approximate solutions, numerical algorithms and ways to improve and operationalize near-real time variants of these algorithms. In other words, besides the tradeoffs dictated by the network flow topology, and functional requirement perspectives, computation constraints are also needed to be considered.

**References and Notes:**

1. S. M. Rinaldi, J. P. Peerenboom, T. K. Kelly, Identifying, understanding, and analyzing critical infrastructure interdependencies. *IEEE Control Systems Magazine*. **21**, 11–25 (2001).

2. M. McNutt, Preparing for the next Katrina. *Science*. **349**, 905–905 (2015).

3. D. Helbing, Globally networked risks and how to respond. *Nature*. **497**, 51–59 (2013).

4. The Global Risks Report 2018. *World Economic Forum*, (available at https://www.weforum.org/reports/the-global-risks-report-2018/).

5. A. A. Ganin *et al.*, Operational resilience: concepts, design and analysis. *Scientific Reports*. **6**, 19540 (2016).

6. J. Gao, B. Barzel, A.-L. Barabási, Universal resilience patterns in complex networks. *Nature*. **530**, 307–312 (2016).

7. K. L. Clark, U. Bhatia, E. A. Kodra, A. R. Ganguly, Resilience of the US National Airspace System Airport Network. *IEEE Transactions on Intelligent Transportation Systems*, 1–10 (2018).

8. S. V. Buldyrev, R. Parshani, G. Paul, H. E. Stanley, S. Havlin, Catastrophic cascade of failures in interdependent networks. *Nature*. **464**, 1025–1028 (2010).

9. M. Ouyang, Z. Wang, Resilience assessment of interdependent infrastructure systems: With a focus on joint restoration modeling and analysis. *Reliability Engineering & System Safety*. **141**, 74–82 (2015).

10. U. Bhatia, D. Kumar, E. Kodra, A. R. Ganguly, Network Science Based Quantification of Resilience Demonstrated on the Indian Railways Network. *PLOS ONE*. **10**, e0141890 (2015).

11. A. Ulusan, O. Ergun, Restoration of services in disrupted infrastructure systems: A network science approach. *PLOS ONE*. **13**, e0192272 (2018).

12. A. A. Ganin *et al.*, Resilience and efficiency in transportation networks. *Science Advances*. **3**, e1701079 (2017).

13. M. Ouyang, L. Dueñas-Osorio, X. Min, A three-stage resilience analysis framework for urban infrastructure systems. *Structural Safety*. **36–37**, 23–31 (2012).

14. U. NIAC, A Framework for Establishing Critical Infrastructure Resilience Goals. *Final Report and Recommendations by the Council, US Department of Homeland Security*. **10** (2010).

15. A. R. Ganguly, U. Bhatia, S. E. Flynn, U. Bhatia, S. E. Flynn, *Critical Infrastructures Resilience : Policy and Engineering Principles* (Routledge, 2018; https://www.taylorfrancis.com/books/9781498758642).

16. E. D. Vugrin, M. A. Turnquist, N. J. K. Brown, Optimal recovery sequencing for enhanced resilience and service restoration in transportation networks. *International Journal of Critical Infrastructures*. **10**, 218–246 (2014).

17. P. M. Murray-Tuite, in *Proceedings of the 38th Conference on Winter Simulation* (Winter Simulation Conference, Monterey, California, 2006; http://dl.acm.org/citation.cfm?id=1218112.1218367), *WSC '06*, pp. 1398–1405.

18. L. Zhang, Y. Wen, M. Jin, The Framework for Calculating the Measure of Resilience for Intermodal Transportation Systems (2009) (available at https://trid.trb.org/view/1089300).

19. R. M. Rauber *et al.*, Finescale Structure of a Snowstorm over the Northeastern United States: A First Look at High-Resolution HIAPER Cloud Radar Observations. *Bull. Amer. Meteor. Soc.* **98**, 253–269 (2016).

20. M. E. J. Newman, Assortative Mixing in Networks. *Phys. Rev. Lett.* **89**, 208701 (2002).

21. V. Latora, M. Marchiori, Is the Boston subway a small-world network? *Physica A: Statistical Mechanics and its Applications*. **314**, 109–113 (2002).

22. L. Lu, M. Zhang, in *Encyclopedia of Systems Biology*, W. Dubitzky, O. Wolkenhauer, K.-H. Cho, H. Yokota, Eds. (Springer New York, New York, NY, 2013; https://doi.org/10.1007/978-1-4419-9863-7_874), pp. 647–648.

23. L. Sela Perelman, W. Abbas, X. Koutsoukos, S. Amin, Sensor placement for fault location identification in water networks: A minimum test cover approach. *Automatica*. **72**, 166–176 (2016).

24. The Cross-Entropy Method for Combinatorial and Continuous Optimization | SpringerLink, (available at https://link.springer.com/article/10.1023/A:1010091220143).

25. J. N. Kapur, H. K. Kesavan, in *Entropy and Energy Dissipation in Water Resources*, V. P. Singh, M. Fiorentino, Eds. (Springer Netherlands, Dordrecht, 1992; https://doi.org/10.1007/978-94-011-2430-0_1), *Water Science and Technology Library*, pp. 3–20.

26. M. Moher, in *Proceedings of GLOBECOM '93. IEEE Global Telecommunications Conference* (1993), pp. 809–813 vol.2.

27. A. Krause, A. Singh, C. Guestrin, Near-Optimal Sensor Placements in Gaussian Processes: Theory, Efficient Algorithms and Empirical Studies. *J. Mach. Learn. Res.* **9**, 235–284 (2008).

28. D. M. Topkis, Minimizing a Submodular Function on a Lattice. *Operations Research*. **26**, 305–321 (1978).

29. R. G. Ingalls, M. D. Rossetti, J. S. Smith, B. A. Peters, *Global Likelihood Optimization Via the Cross-Entropy Method with an Application to Mixture Models*.

30. S. Kullback, R. A. Leibler, On Information and Sufficiency. *Ann. Math. Statist.* **22**, 79–86 (1951).

**Acknowledgments: Funding:** This work was supported in part by the Civil and Environmental Engineering Department, Sustainability and Data Sciences Laboratory, Northeastern University. The work of U. Bhatia and A. R. Ganguly were supported by four National Science Foundation Projects, including NSF BIG DATA under Grant 1447587, NSF Expedition in Computing under Grant 1029711, NSF CyberSEES under Grant 1442728, and NSF CRISP type II under Grant 1735505. **Author contributions:** UB, LS and ARG designed the experiments, UB and LS performed the analysis. All authors contributed equally in writing the manuscript. **Competing interests**: Authors declare no competing interests.

**Data and materials availability:** Data and materials availability is discussed in manuscript and SI and cleaned datasets are available upon request to authors.

**Supplementary Materials:**

Materials and Methods

Figures S1

Tables S1-S3

References (*23-30*)

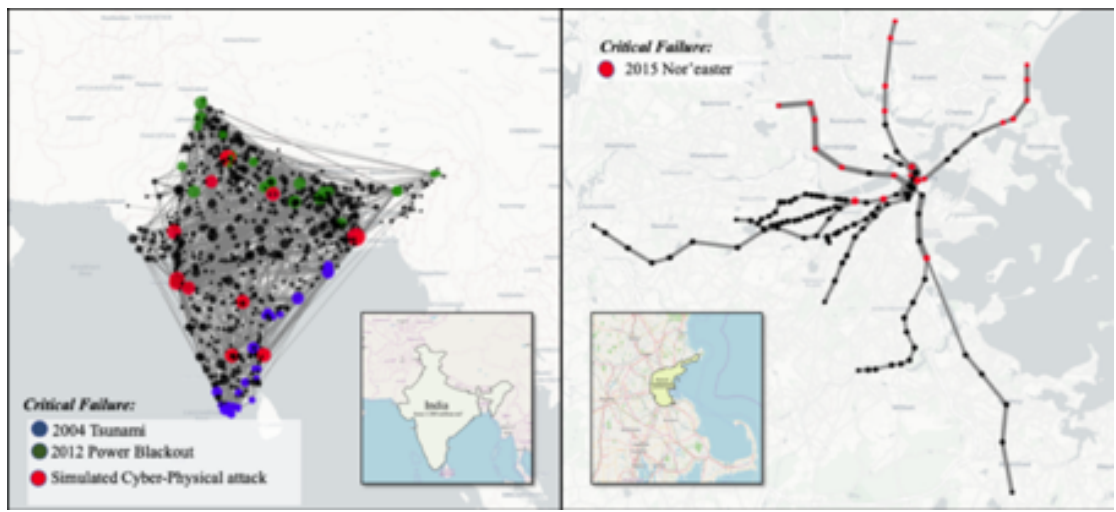

**Fig. 1. Network visualization of Transportation Systems.** Figure 1(A): Network visualization of Indian Railways Network (IRN). Node size is proportional to the number of connections of each network. Critical failure nodes impacted from simulated cyber-physical hazards, 2004 Indian Ocean Tsunami, and 2012 India Blackouts are shown in red, blue, and green respectively. Figure 1(B): Same as Figure 1(B) but for Massachusetts Bay Transportation Authority (MBTA) Mass Transit System. Nodes impacted by 2013 snowstorms are shown in red. Geographical maps in inset highlights the different scales at which these two transportation networks operate.

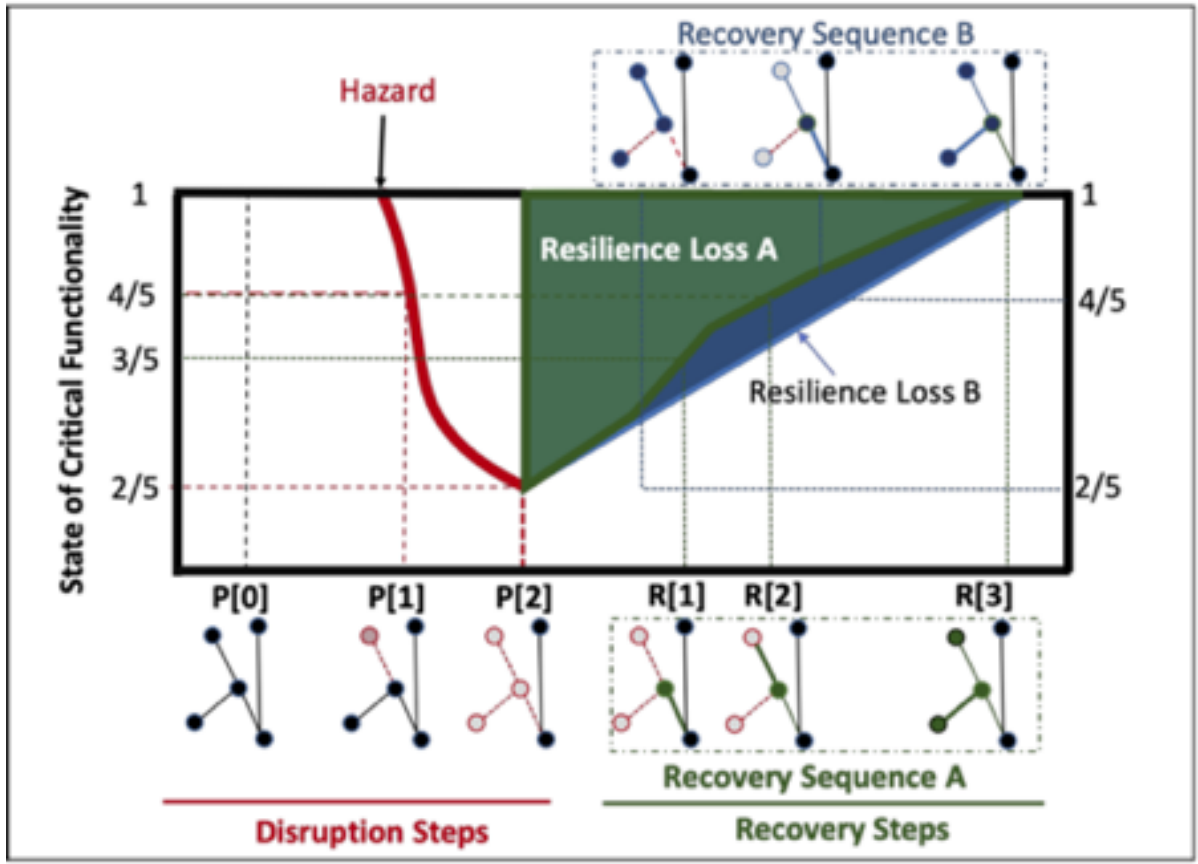

**Fig. 2. Disruption and recovery process in representative network.** For the representative network with 5 nodes and 4 edges we measure SCF at time t as ratio of number of nodes at time=t with respect to original number of nodes at T=0. At time P [0], SCF = 1 (pre-hazard). Edge shown in red and marked by arrow is selected for removal at t = P[1], resulting in isolation of the node shown in red, which sets SCF to 4/5. At P[2], another edge is removed, which further shrunk SCF to 2/5. After network is disrupted, recovery process is initiated. With three edges removed, edge restoration can be determined in 6 ways. Out of 6 possible way, we consider recovery sequence A and B. In recovery sequence A, when green edge is restored at time = R [1], SCF changes from 2/4 to 3/5. The process of restoring one edge at a time is repeated until the network has SCF =1 (pre=hazard state). Efficiency of strategy A is measured in terms of resilience loss, which is area bounded between Y-axis and recovery curve traced by SCF. Similarly, recovery sequence B traces its own recovery curve. Strategy with least resilience loss is comparatively more efficient.

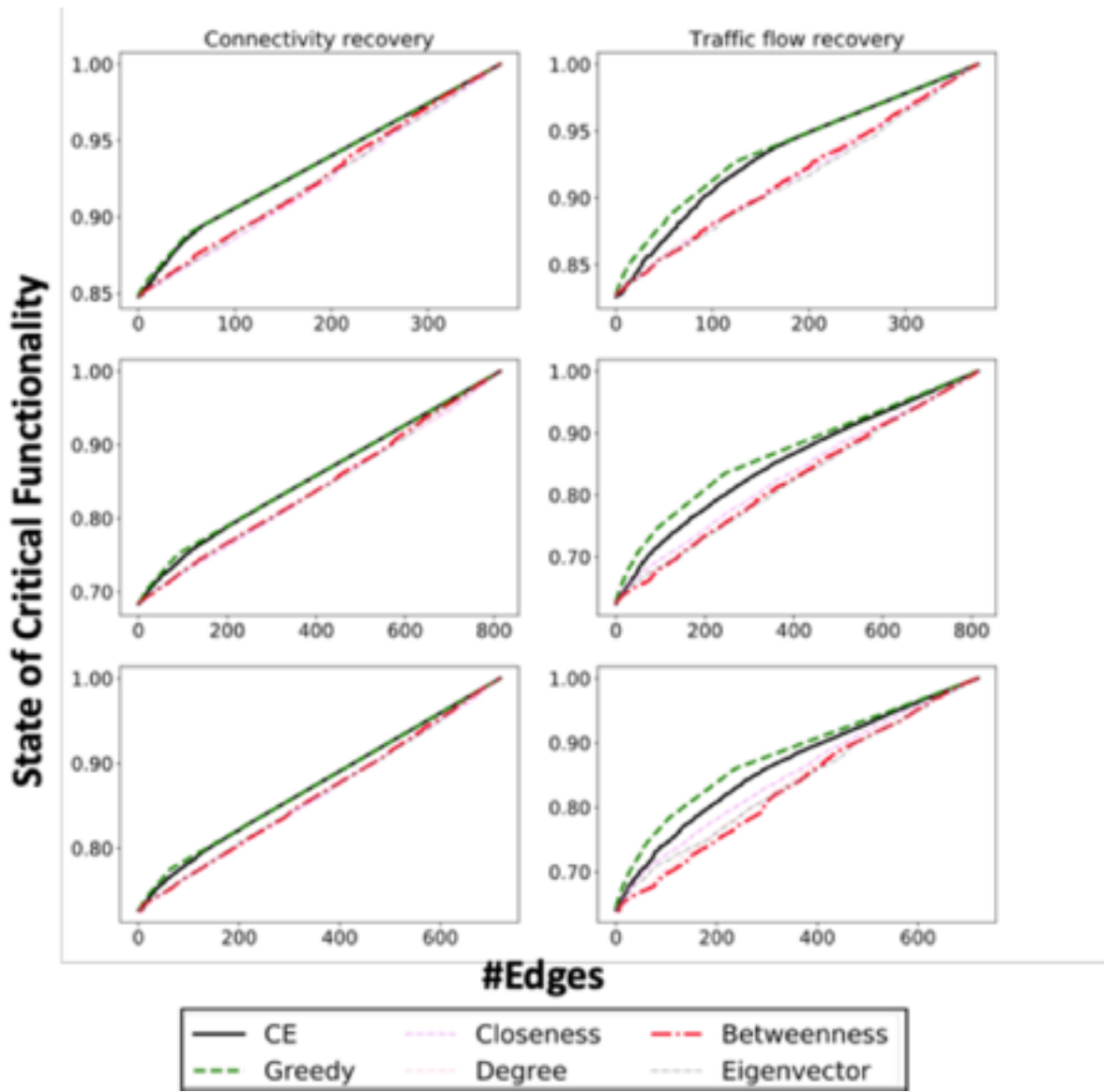

**Fig. 3: Recovery of Indian Railways Network after three hazards.** Each row corresponds to a specific hazard and column corresponds to the performance measure used to calculate State of Critical Functionality. Five approaches (2 optimization based, and 3 network science based) are used to generate edge prioritization sequences for perturbed sequences. Out of five approaches, the two optimization approaches and one network-based approach (edge betweenness centrality) is shown in bold. Top row: simulation inspired by 2004 Indian Ocean tsunami; middle row: scenario based on a cascade from the power grid; and bottom row: a cyber or cyber-physical attack scenario. In all three cases and for both performance measures, optimization-based recovery approaches are more efficient than network centrality-based approaches, on an average, by 9% in terms of resilience loss (area between Y-axis and recovery curve). Table 1 summarizes the resilience loss calculated for the three hazards and two performance measures

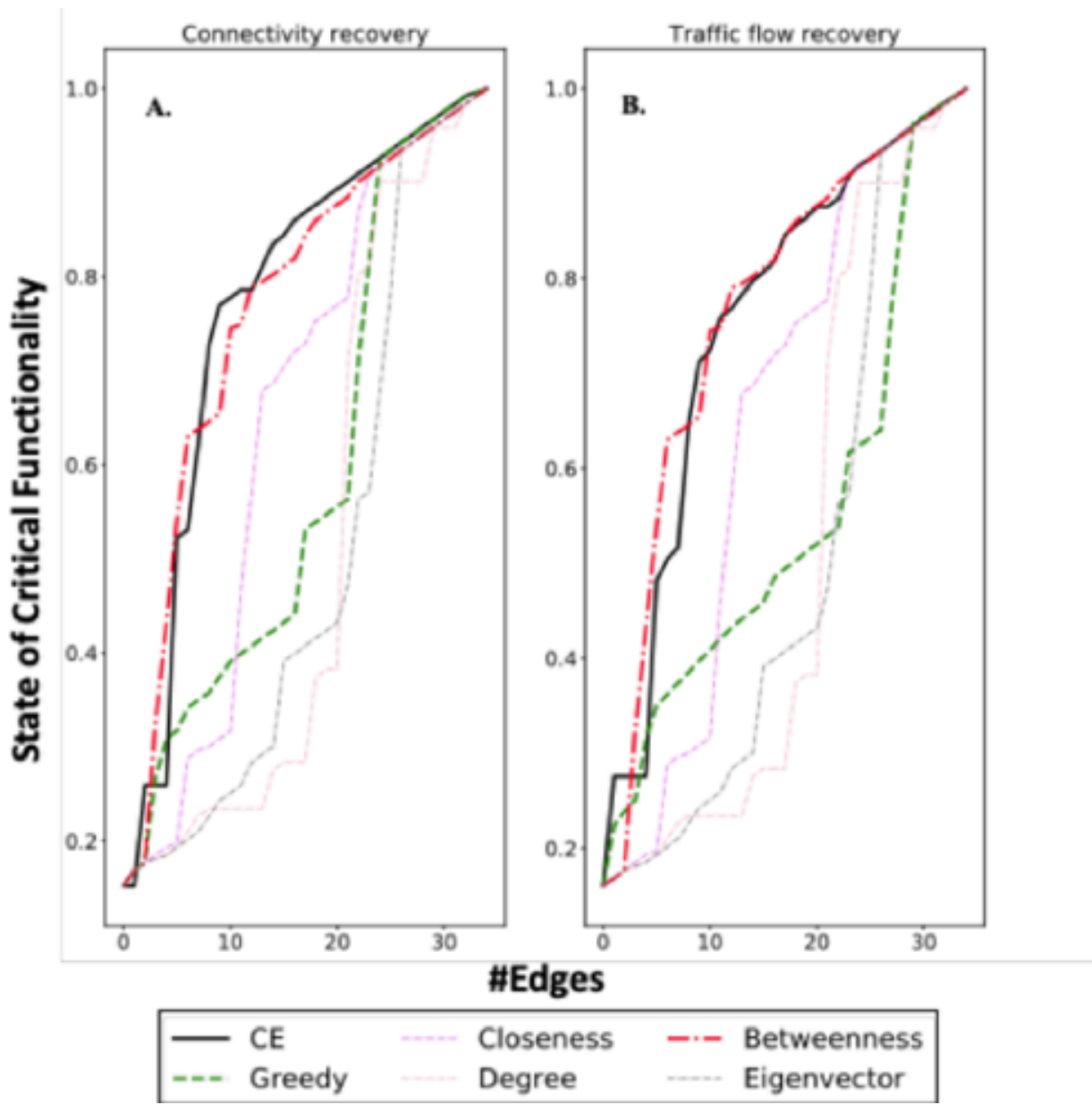

**Fig. 4. Recovery of MBTA after 2015 Nor'easter.** Five approaches (2 optimization based, and 3 network science based) are used to generate edge prioritization sequences for perturbed sequences. Out of five approaches, the two optimization approaches and one network-based approach (edge betweenness centrality) is shown in bold. **A:** Recovery of SCF in terms of network connectivity shows that network-based edge betweenness perform as good as CE optimization, and both these approaches outperform GA by nearly 41% in terms of efficiency **B.** Similar insights are obtained when traffic volume is used to measure SCF. Table 2 summarizes the resilience loss calculated for two performance measures.

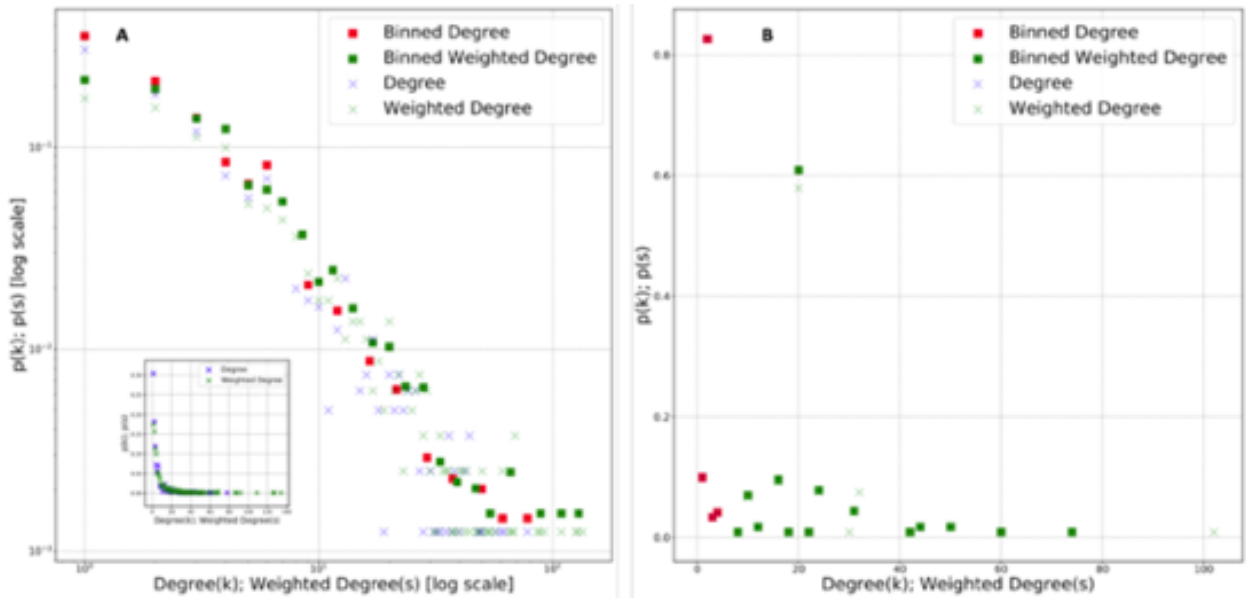

**Fig. 5. Topology of the two networks. (A)** Probability distribution of node degree and weighted degree of stations in Indian Railways Network (IRN) on a log- log scale. The distributions follow power law model, wherein most stations have a small number of connections, with the exception of a few stations having exceptionally large degrees in comparison to average degree Figure in inset shows degree distribution on linear scales. To reduce the effects of discontinuity in degree distribution, a number of more or less continuous values are grouped together in "bins'' to generated binned degree and weighted degree. Figure in inset shows degree and weighted degree distribution on linear scale. **(B)** same as right but for MBTA on linear scale. MBTA exhibits a tree-like topology with 80% nodes having degree between 1 and 3. However, weighted degree is comparatively more scattered in comparison to degree based on number of connections. Since no exponential decay or power-law behavior is observed in degree and weighted degree distribution, linear scale is used.

**Table 1. Recovery scores for Indian Railways network:** Resilience loss, measured as area between Y axis and State of resilience curve traced by each strategy curve during recovery process. Lower area means more efficient recovery strategy. Most efficient recovery strategies are shown in **red.**

| Hazards on Network: IRN → | Tsunami | | Grid | | Cyber-physical attack | |
|---|---|---|---|---|---|---|
| Edges recovered→ | 376 | | 815 | | 720 | |
| Recovery algorithm↓ | GC | OD | GC | OD | GC | OD |
| Greedy Optimization | **26.6** | **25.15** | **116.92** | **108.73** | **91.27** | **87.7** |
| Closeness | 29.79 | 34.47 | 128.65 | 150.98 | 99.75 | 128.87 |
| Degree | 29.24 | 34.18 | 128.69 | 247.59 | 100.06 | 126.24 |
| Betweenness | 29.2 | 33.93 | 127.65 | 147.76 | 99.63 | 126.34 |
| Eigenvector | 29.88 | 34.63 | 128.5 | 150.52 | 100.13 | 130.42 |
| Cross Entropy | **26.75** | **26.6** | **118.01** | **120.93** | **92.02** | **98.8** |

**Table 2. Recovery scores for Boston's MBTA.** Same as Table 1 but for MBTA. Most efficient recovery strategies are shown in red.

| Network: MBTA | 2015 Northeaster', Boston | |
|---|---|---|
| Edges recovered→ Recovery algorithm↓ | 35 | |
| Greedy Optimization | 17.26 | 15.95 |
| Closeness | 11.16 | 11.26 |
| Degree | 14.30 | 13.85 |
| Betweenness | **10.18** | **9.77** |
| Eigenvector | 16.8 | 17.84 |
| Cross Entropy | **10.31** | **9.74** |

# Supplementary Information

**This PDF file includes:**

Supplementary Information text
Figs. S1
Tables S1 to S3

**Supplementary Information Text**

**Materials and Methods**

*Data*

We generate models of two topologically distinct transportation systems operating at disparate scales. Specifically, we model Indian Railways Network (IRN) and Boston's Mass Transit System (MBTA) as networks. In both networks, stations are modeled as nodes. For IRN, a pair of nodes is considered being connected by an edge if there is at least one origin and destination train between the pair. for MBTA, two nodes are considered being connected if a pair of station is connected by a direct train. We use two different network representations because both Origin Destination (OD) and traffic flow data is frequently used in transportation for efficiency and resilience assessment studies depending upon the data availability (*16–18*). Here, we analyze origin-destination data of passenger-carrying trains on the IRN. The network is constructed using publicly available data, which was cleaned and appropriately formatted prior to analysis. We considered stations with at least one originating or terminating train, comprising a total of 809 stations with 7066 trains. We construct MBTA network by analyzing open-source map available at the operator's website. Traffic flow data is obtained from https://www.mbta.com/schedules/. The resulting MBTA network has 121 nodes and 176 edges. We note that traffic flow is a dynamic attribute of any transit network and it depends on multiple factors such as average delays, weather conditions, physical state of stations and tracks. However, these datasets are not publicly and widely available for both IRN and MBTA. Hence, we relied on open-source datasets for this study.

*Methods*

We built on Network Centrality (NC) measures to generate prioritization sequences for perturbed systems. Different centrality indices measure distinctive aspects related to a position of nodes within a network. For example, Betweenness Centrality describes importance of node as a connector between different parts of the network. Alternatively, Closeness Centrality measures the proximity of a node to all other nodes. Nodes with high closeness centrality values can rapidly affect other nodes and vice versa. Degree centrality is a measure of the number of connections originating (or terminating at a node), whereas eigenvector centrality is measure of influence of the node. A high eigenvector centrality means that node is connected to many more nodes which have a high degree. Similarly, edge betweenness centrality is defined as a number of the shortest paths that go through an edge in a network.

We consider two performance metrics, where the resilience is a function of (a) network connectivity and (b) the origin-destination demand that can be satisfied. For (a), we consider topology-based centrality measure whereas for (b) weighted version of NC is used, where weight of an edge represents the number of trains scheduled to operate on a given edge (24). Specifically, we use weighted degree (sum of weights incident upon a node), weighted closeness centrality (measuring proximity of node while using inverse of weights to find least costly path among all nodes), and weighted edge betweenness centrality based on the algorithm proposed by Newman (25).

Intuitively, an edge connecting two important nodes should act like a bridge between two parts of a network. Hence, removal(restoration) of such an edge could result in faster destruction (recovery) of the system. Hence, in addition to edge betweenness centrality, we also account for importance of a pair of nodes between which an edge is positioned. We compute averages of closeness, average eigenvector, and average degree centrality for each pair of nodes in the network and centrality-based scores are used as measures to establish edge prioritization sequence. The higher the centrality-based score, the higher prioritization to be assigned to an edge during recovery process.

We design two optimization approaches to solve the network recovery problem, specifically, the greedy approach (GA) and the Cross Entropy (CE) stochastic optimization On the contrary, the GA has the disadvantages of being trapped by local optimums and thus not reaching globally good solutions. The CE relies on large number of simulations, which typically makes it computationally intense especially for large systems. The main drawback of both approaches is that they do not provide performance guarantees on the quality of the solution. However, they have been highly successful for applications in variety of computer science and optimization problems and empirically result in good quality solutions (*25*, *26*).

To formulate the recovery problem, we define the recovered edges as the decision variables, where if edge *i* recovered at time t and 0 otherwise. Our goal is to achieve a fast recovery to the pre-disaster functionality, i.e. minimize resilience loss. Thus, we define the objective function as the area under the resilience curve that we wish to minimize. We consider two performance metrics, where the resilience is a function of network connectivity and the origin-destination demand that can be satisfied. The outcome of the optimization approach, either GA or CE, would provide the sequence of the edges that should be restored to achieve the minimum resilience loss.

Intuitively, GA approach is to iteratively select the solution (edge) that contributes the most to the network functionality at the current stage, until all edges are restored, and the network regains its full functionality. The main steps of the greedy approach are outlined in Table S3 and full description can be found in (*24*). The GA has been previously widely studied in the context of submodular function optimization and combinatorial optimization in a wide range of applications, e.g. sensor placement and scheduling (*27*). In the case where the objective function is monotone submodular (*28*), the greedy approximation can provide performance guarantees. For the recovery problem, this means that as the number of recovered edges increases the marginal value of system functionality decreases. This assumption does not hold in our case and we cannot provide performance guarantees. However, in practice, the GA still shows better performance than the theoretical guarantees.

The recovery problem was also solved using Cross Entropy (CE) method for combinatorial optimization (*24*, *29*). The cross entropy (CE) algorithm is a heuristic search technique, which utilizes probabilities of outcomes of the decision variables instead of the actual values. Intuitively, the CE associates probabilities with the decision variables and the goal is to find the optimal probabilities distributions for the decision variables. For example, $p(x_{i,t})$ represents the probability of recovering edge *i* at time *t*, starting with a random probability, the CE will converge when the probabilities are close to $p(x_{i,t})$ representing that edge *i* is recovered at time t and 0 otherwise. To find the optimal probabilities, the CE algorithm relies on a two-stage iterative process: (a) generating potential solutions from the sampling probability and (b) updating of the parameters of the sampling probabilities to find better solutions by minimizing the Kullback-Leibler distance (*30*) between the sampling probability and the theoretical optimal probability. The CE algorithm has three parameters, which control its performance including: sample size, which defines the number of samples in each iteration; the elite sample percentage, which defines the best set of solutions used for updating the parameters of the sampling probability; and a smoothing parameter, which prevents from premature convergence. The main steps of the CE approach are outlined in Table S3 and a detailed description can be found in (*24*).

<insert page break then Fig. S1 here>

**Fig. S1. Recovery curves for MBTA for different choice of parameters :** To ensure that our results are robust to the choice of parameters, we perform CE optimization using 3 different choices of parameters and we observe that efficiency scores thus obtained are insensitive to parameter choice for both objective functions. A. SCF in terms of network connectivity B. SCF in terms of traffic flow.

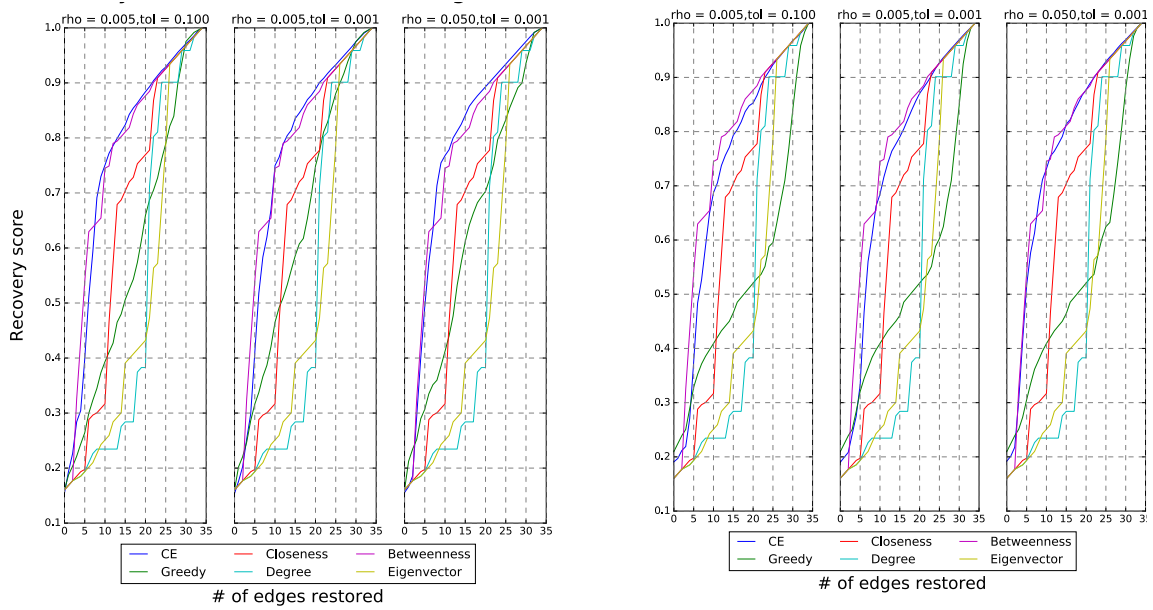

Table S1 **List of the stations removed for each of the three hazards for Indian Railways Network**

| Tsunami | Power Failure | Cyber-Physical |
|---|---|---|
| Bhubaneswar | Ajmer | Ahmedabad |
| Chennai Central | Ambala Cant | Bangalore City |
| Chennai Egmore | Amritsar | Chennai Central |
| Cuttack | Anand Vihar | Chennai Egmore |
| Ernakulam | Asansol | Delhi |
| Gudivada | Bareilly | H Nizamuddin |
| Guntur | Bareilly City | Howrah Jn |
| Kanyakumari | Bikaner | Hyderabad Decan |
| Kochuveli | Danapur | Jaipur |
| Kollam Jn | Darbhanga | Kacheguda |
| Machilipatnam | Delhi | Kolkata |
| Madurai | Delhi S Rohilla | Lokmanyatilak |
| Mayiladuthurai | Dibrugarh | Mumbai |
| Nagercoil | Dibrugarh Town | New Delhi |
| Narasapur | Firozpur Cant | Pune |
| Puducherry | Ghaziabad | Sealdah |
| Puri | Gorakhpur | Secunderabad |
| Rameswaram | Guwahati | Shalimar |
| Sengottai | H Nizamuddin | Yesvantpur |
| Tiruchendur | Howrah | |
| Tiruchirapalli | Jaipur | |
| Tirunelveli | Jammu Tawi | |
| Tirupati | Kamakhya | |
| Trivandrum | Kanpur Anwrganj | |
| Tuticorin | Kanpur Central | |
| Vijayawada | Kolkata | |
| Villupuram | Lal Kuan | |
| Visakhapatnam | Lucknow | |
| | Muzaffarpur | |
| | New Delhi | |
| | New Tinsukia | |
| | Palwal | |
| | Patna | |
| | Pratapnagar | |
| | Puri | |
| | Sealdah | |
| | Tinsukia | |
| | Varanasi | |

**Table S2: List of the stations removed from Boston's Mass Transit System during simulation of Nor'easter of 2015**

| |
|---|
| Oak Grove |
| Malden Center |
| Wellington |
| Assembly |
| North Station |
| State |
| Copley |
| Kenmore |
| Wonderland |
| Revere Beach |
| Beachmont |
| Suffolk Downs |
| Orient Heights |
| Aquarium |
| Bowdoin |
| Alewife |
| Davis |
| Porter |
| Harvard |
| Central |
| Charles MGH |
| UMASS |

**Table S3: Summary of optimization-based recovery algorithms used for network recovery.**

| Algorithm 1: Greedy approach | Algorithm 2: Cross Entropy (CE) |
|---|---|
| 1. **Input:** a function which evaluates the value of a solution; a candidate set from which a solution is created<br>2. **Select:** the element that results in the maximum improvement in system functionality<br>3. **Repeat:** until all resources are exhausted or no further improvement can be made | 1. **Initialize:** the parameters of the sampling distribution<br>2. **Generate:** random samples from the sampling distribution<br>3. **Update:** the parameters of the sampling distribution by minimizing the Kullback-Leibler distance measure.<br>4. **Repeat:** from step 2 until some stopping criteria is met |